\documentclass[a4paper]{jpconf}
\usepackage{graphicx}
\begin{document}
\title{On (Schr\"{o}dinger's) quest for new physics for life}

\author{Alfredo Iorio}

\address{Institute of Particle and Nuclear Physics, Charles University of Prague, V Hole\v{s}ovickach 2, 18000 Prague 8, Czech Republic}

\ead{Alfredo.Iorio@mff.cuni.cz}

\begin{abstract}
Two recent investigations are reviewed: quantum effects for DNA aggregates and scars formation on virus capsids. The possibility that scars could explain certain data recently obtained by Sundquist's group in electron cryotomography of immature HIV-1 virions is also briefly addressed. Furthermore, a bottom-up reflection is presented on the need to invent new physics to pave the way to a rigorous physical theory of biological phenomena. Our experience in the two researches presented here and our personal interpretation of Schr\"{o}dinger's vision are behind the latter request.
\end{abstract}

\section{Introduction}

It seems to us that the frontier of physics is rapidly moving to the biological side not only in the obvious way that
the highly detailed knowledge of the biological data requires increasingly sophisticated physics for their
explanation, but also (and even more intriguingly for the theoretical physicist's mind) in the sense that the time might be mature for
a profound reformulation of biology that makes it a science similar to physics: few principles, mathematical
derivations, explanation and prediction of phenomena (and back). Nonetheless, the problem is very difficult if faced with the proper rigor and
we have no answer to propose, a part from the conviction that some important changes are necessary on the physical approach
to the problems of biology and that a ``physics tailored for life phenomena'' needs to be invented. On this we comment only in the last section where
we arrive by taking a feet-on-the-ground road that makes two stops prior to getting there: the study of (relativistic) quantum effects
in DNA condensation \cite{issletter} and the role of certain topological defects (scars, see, e.g., \cite{Iorio:2006ur}) for virus capsids shape-changes \cite{cejb2008}. This way we have the chance to face important biological problems on the table and to try to learn on the battle-field what is that is missing from the big picture.

\section{Quantum fields holding together DNA aggregates}

\subsection{The story}

DNA molecules in aqueous solution ionize and become highly charged anions which strongly repel \cite{kholodenko}-\cite{qiu} (for reviews see, e.g., \cite{gelbart2000, KLreview}). When specific cations are added (i) DNA attracts and {\it binds} them to make a new structure, the DNA-cations complex, and (ii) when about 90 per cent of the DNA negative charge is screened the {\it like-sign} DNA-cations strands {\it attract and collapse} (see, e.g., \cite{gelbart2000, KLreview, Levin:2002gj}) to form finite-size aggregates whose shape is either rod-like or spheroidal or, most commonly, toroidal \cite{Hud}.

Long ago Oosawa and Manning (OM) explained the counterions condensation as a phase transition within the classical Poisson-Boltzmann (PB) mean field
electrostatic theory \cite{oosawa}. Despite important advancements, the second part of the puzzle still presents many open questions: (i) it is not understood why the aggregate does not grow forever; (ii) there is no general consensus on the necessity to go beyond the PB classical theory, as proposed in \cite{oosawabook} (see also \cite{groenbech}) and as opposed in \cite{kornyshev1998} (see also \cite{KLreview, cherstvy2002-2005});
(iii) whichever approach is used (the zero frequency Casimir/van der Waals interaction \cite{parsegian1970} (see also \cite{richmond}); the ``electrostatic zipper'' model of \cite{kornyshev1998}; correlations of thermal Gaussian fluctuations of the number density of counterions \cite{groenbech, oosawabook}; the Wigner crystal approach \cite{grason, rouzina}) the paradigm is that the interactions are {\it classical} \cite{gelbart2000}. The lack of appreciation of quantum effects for this phenomenon was probably due to the lack of calculations based on the appropriate {\it codimension two} (lines in three dimensions) Casimir-like technique\footnote{The work in \cite{parsegian1970, parsegian1998} is based on the Lifshitz computation for the Casimir effect in presence of dielectric media \cite{Schwinger} (see also \cite{milonni} and references therein) i.e. a codimension one (surfaces in three dimensions) calculation.} \cite{Scardicchio:2005hh} and with cations not directly participating in the interaction but playing the passive role of screening the electrostatic repulsion.

\subsection{A new model based on relativistic quantum fields}

In \cite{issletter} we singled-out the zero-point interaction due to the disturbances induced in the \textit{quantum electric vacuum} by the presence of the (nearly) neutral DNA-cations complexes. We modeled the $N$ anions (the DNA strands) as infinite lines (finite length effects are unimportant) all parallel to the $z$-axis and located at $\vec{l}_i$ in the $x-y$ plane with certain coefficients $\nu_i (z_i)$ carrying information on the charge structure of the DNA and taken to be $\nu_i (z_i) = \nu =$~constant, $\forall i = 1, ..., N$. The cations screen the total charge and set the length scale. The effective model for the electric potential $\Phi (\vec{x})$ in this set-up is given by \cite{issletter}
\begin{equation}\label{DHmod}
\left[ - \partial^2_z - \nabla_\bot^2 + \mu^2 + \lambda \sum_{i=1}^N \delta^{(2)} (\vec{x}_\bot - \vec{l}_i) \right] \Phi(\vec{x}) = J \;,
\end{equation}
which is a {\it modified} Debye-H\"{u}ckel (DH) equation, with $\mu^2 = k^2 \kappa^2$, $k$ the cations valency, $\kappa^{-1} = (\epsilon k_B T/(8 \pi e^2 n_0))^{1/2}$ the Debye screening length, $\epsilon$ the dielectric constant of water, $n_0$ fixes the zero of the potential, $\lambda = 4 \pi \nu |q|^2 / \epsilon k_B T$, $J =  - (1 / \epsilon) 4 \pi |q| \nu \sum_{i=1}^N \delta^{(2)} (\vec{x}_\bot - \vec{l}_i)$, $T = 300$K, and with Eq.(\ref{DHmod}) the limit for weak $\Phi$ of a PB equation modified to include the DNA strands in the charge (Boltzmann) distribution \cite{issletter}
\begin{equation}
\rho_{\rm DNA} (\vec{x}, T) = - n^0_{\rm DNA} (\vec{x}) |q| \exp
\left(  \frac{|q| \Phi (\vec{x})}{k_B T} \right)
\end{equation}
where $q < 0$ is the charge of the DNA strand with $n^0_{\rm DNA} (\vec{x}) = \sum_{i=1}^N \nu_i (z_i) \delta^2 (\vec{x}_\bot - \vec{l}_i)$ and $N$ the number of strands.

Eq.(\ref{DHmod}) is only a starting point, as our concern are time dependent fluctuations, $\Phi (\vec{x}) \to \Phi (\vec{x}) + \phi (\vec{x}, t)$, that we treat as quantum by considering the associated action (we set $\hbar = c = 1$, $c$ speed of light in the medium)
\begin{equation}
\bar{{\cal A}} (\phi) = \int d^4 x \frac{1}{2} \phi \left( - \partial_t^2 - \partial^2_z - \nabla_\bot^2 + \mu^2 + \lambda \sum_{i=1}^N \delta^{(2)} (\vec{x}_\bot - \vec{l}_i) \right) \phi \;,
\end{equation}
which we infer from the action for $\Phi$ that gives Eq.(\ref{DHmod}) as equation of motion, ${\cal A} (\Phi) = \int d^4 x  \left( \frac{1}{2} \Phi [- \partial^2_z - \nabla_\bot^2 + \mu^2 + \lambda \sum_{i=1}^N \delta^{(2)} (\vec{x}_\bot - \vec{l}_i)] \Phi + J \Phi \right)$,
to be of the order of $\hbar$. Note that in $\bar{{\cal A}} (\phi)$ the term with the coupling to $J$ is zero because
$\int d^4 x J \phi = \int d^3 x J \int_0^\tau dt \phi = 0$.

These fluctuations are disturbances of the quantum vacuum induced by the delta-functions and, since they are the electric field, they travel at the speed of light in the medium. We considered the generating functional $Z[\Phi, \phi]  =  \int [D\Phi] e^{i {\cal A} (\Phi)} \int [D\phi] e^{i \bar{{\cal A}} (\phi)} = \int [D\Phi] e^{- ( {\cal A} (\Phi) + {\rm corrections})}$ where we Wick rotate on the time direction $t \to i t$, and by identifying the effective action as ${\cal A}_{\rm eff} (\Phi) = {\cal A} (\Phi) + {\cal E} \tau$ one sees that \cite{issletter}
\begin{equation}
{\cal E} = \frac{1}{2} \int_{-\infty}^{+\infty} \frac{d p}{2 \pi} \int_0^{+\infty} dE \rho (E) \sqrt{E + p^2} \sim \frac{1}{2} \sum_k \omega_k
\end{equation}
where $\rho(E)$ is the density of states, i.e. $\cal E$ is the zero point ``Casimir-like'' energy of the system, that is found to be \cite{issletter}, \cite{Scardicchio:2005hh} (see also \cite{Jaffe:2005wg})
\begin{equation}\label{Elndet}
{\cal E} = \frac{\hbar c}{8 \pi} \int_0^\infty dE \ln \left[ \det \left( \delta_{i j} -  \frac{K_0 (\sqrt{E
+ \mu^2} \; l_{i j})}{\ln(\sqrt{E + \mu^2} / M)} (1 - \delta_{i j})
\right) \right] \;,
\end{equation}
where we reintroduced $\hbar$ and $c$, $K_0 (x)$ is the modified Bessel function of the second kind of order zero, $\mu$ is the scale parameter introduced earlier, $l_{i j} = |\vec{l}_i - \vec{l}_j|$ are the relative distances and $M$ is a further mass scale parameter originated by the codimension two (for stability it satisfies $M < \mu$). We are not considering charge density or positional fluctuations of counterions, hence we do not have slow-moving fluctuations, as, e.g., the authors of \cite{parsegian1998} that had to consider only zero frequency (i.e. classical) contribution to the usual Casimir/van der Waals effect. We have here a universal mechanism of interaction that (i) only becomes important when the OM condensation has taken place, (ii) whose temperature dependence is via the length scale $\mu$ and (iii) with a free (but highly constrained in units of $\mu$) parameter $M$.

\begin{figure}
 \centering
  \includegraphics[height=.2\textheight]{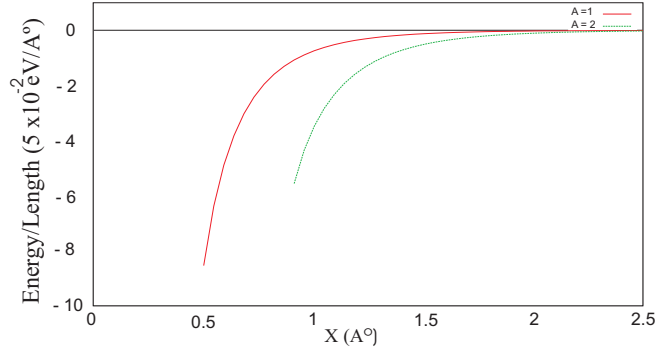}
  \caption{Two-Body interaction energy/length ${\cal E}(x)$ against lattice size $x$ for $a=1$ (upper curve) and $a=2$ (lower curve)}
\label{2_String_Energy}
\end{figure}

\subsection{New results}

\begin{figure}
 \centering
  \includegraphics[height=.2\textheight]{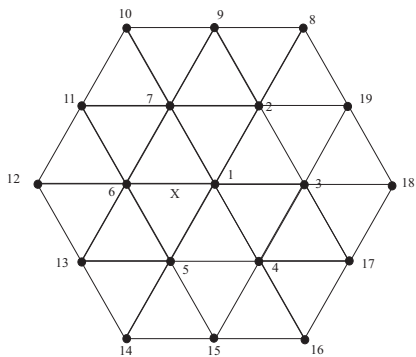}
  \caption{Configuration used to compute the interaction of 19 strands.}
\label{config19}
\end{figure}

The two-body interaction indeed is attractive as shown in Fig.~\ref{2_String_Energy}. Our focus needs be on configurations as close as possible to real cases \cite{Hud}, such as that shown in Fig.~\ref{config19}, where the relative distances are $l_{i j} = c_{i j } x$ with $x$ the basic lattice distance and $c_{i j}$ taking the symmetry of the given arrangement into account. The units are $\mu^{-1} \sim {\cal O} (10)$ \AA, for distances, and ${\cal E} \sim 5 \times 10^{-2}$~eV/\AA, for lineal energy density. $M = e^{-1/a}$ (we choose $a = 1$ and $a = 2$). Thermal fluctuations, as computed, e.g., in \cite{groenbech}, give for the two-body interaction a maximum value of ${\cal E} \sim 5 \times 10^{-3}$~eV/\AA~at 10 \AA. At this distance our two-body interaction is $5 \times 10^{-2}$~eV/\AA~(for $a=1$) or $2 \times 10^{-1}$~eV/\AA~(for $a=2$), i.e. between one and two orders of magnitude stronger. For the many-body case, the case of importance for the aggregates, this factor grows enormously \cite{issletter} but we cannot trust our approximations for $x$ too close to the singular value of the logarithm. It is clear, though, that at the distances of relevance this quantum energy is stronger (or much stronger) than thermal energy. Thus we can conclude that it is relativistic quantum field zero-point energy that holds together DNA aggregates! This is nice but...

\subsection{Effects of the lack of a general theory of biological physics}\label{dnaquestions}
... important pieces are missing here due to the lack of a proper ``general frame'': (i) where is, in the physics of the problem, that those filaments are the carriers of genetic information? (ii) is the description of this phenomenon complicated because we are not using the most appropriate approach? and is that meaning that we need new variables or a different coarse-graining or that we need to take into account that DNA compaction is just part of a series of ``moves`` made with a ``purpose'' (e.g., to pack the DNA in the bacteriophage's head or into the sperm cell and that is, in turn, just one step of virus assembly or of fecundation, and the latter one step in the ``life'' cycle of the virus or in the reproduction of eukaryotic cell organisms, and so on to higher and higher levels of complexity) (iii) how can we translate into physical language the ``purpose``, or, in other words, the clear fact that all these moves are done by ``using the laws of physics for a goal''? is it matter of changing the definition of physical rules or what? finally, (iv) is quantum physics used to full strength, i.e. as pivotal for life itself \cite{schroedinger} (see also \cite{penrose})?

\section{Scars on viruses: the general conjecture and the case of immature HIV-1}
\subsection{The two stories: virus structure and Thomson problem}
A virus is a piece of DNA or RNA surrounded by a protein coat, the capsid, sometimes cased into a lipidic membrane, the envelop \cite{general}. Capsid's shapes can be: helical, icosahedral or complex (sphero-cylindrical, conical, tubular or more complicated shapes) \cite{general}. Viruses may change their shape (polymorphism) as an important step in their ``life''-cycle. For instance, HIV-1 is only infective when its capsid has changed from spherical to conical (maturation) \cite{sundquistreview2008}.

The theory of icosahedral capsids was proposed by Crick and Watson \cite{crickwatson} and later established by Caspar and Klug (CK) \cite{casparklug}:  those capsids all have $60 T$ proteins arranged into 12 pentamers and $10 (T -1)$ hexamers, where $T = n^2 + m^2 + n m = 1,3,4,7,...$. These numbers descend from the Poincar\`e-Euler theorem\footnote{$N_p$ is the number of $p$-gons used to tile a surface, e.g., $N_5$ pentagons, $N_6$ hexagons, etc. The resulting polyhedron $P$ has $V_P = 1/3 \sum_N N_p p$ vertices, $E_P = 1/2 \sum_N N_p p$ edges, and $F_P = \sum_N N_p$ faces, giving for the Euler characteristic $\chi = V_P - E_P + F_P$ the expression in Eq.~(\ref{ngons}).}
\begin{equation}\label{ngons}
  \sum_N (6 - p) N_p = 6 \chi \;,
\end{equation}
applied to the sphere, $\chi = 2$, and from the CK ``quasi-equivalence'' principle \cite{casparklug}, see also \cite{cejb2008}. The CK theory is nowadays an established paradigm among virologists \cite{general} and various modifications/generalizations have been proposed by physicists and mathematicians \cite{cejb2008, twarock, bruinsmazandi, zandi, nguyen} (for a review see \cite{zlotnick}), but none made its way to virology textbooks.

An intimately related physical set-up is that of $N$ electrons on the surface of a sphere. To find their minimum energy configurations means to solve
the so-called ``Thomson problem'' \cite{thomson}, unsolved in general. For $N < {\cal O} (500)$ (and of the form $N = 10 T + 2$) the icosahedron is the solution, as shown in \cite{altschuler}, a work inspired by the CK theory of virus capsids. For bigger $N$ configurations that present ``scars'' are the solution as shown in \cite{perez-garrido} and for spherical elastic materials (non-Coulomb potentials) in \cite{bnt}. A scar is a lineal pattern of pentagons-heptagons that initiate at the pentagonal vertices of the icosahedron, whose length (i.e. number of pentagons and heptagons in it) and geometrical arrangement depend on $N$ (see \cite{Iorio:2006ur} for an approach based on spontaneous symmetry breaking of the icosahedral group). Scars have been experimentally found (in spherical crystals of mutually repelling polystyrene beads self-assembled on water droplets in oil \cite{realscars}) but still need a deep understanding in terms of phase transitions. Nonetheless, it is easy to see that they are allowed by the theorem (\ref{ngons}) written as
\begin{equation}\label{sphere}
    (6 - 5) \;  N_5 + (6 - 6) \; N_6 + (6 - 7) \; N_7 = 12 \;.
\end{equation}
This means that $N_6$ can be arbitrary (hence also the required $N_6 = 10 (T-1)$) and $N_5 - N_7 = 12$, hence, starting off with the 12 pentagons of the icosahedron one can add {\it pairs} pentagon-heptagon, but not a pentagon or a heptagon separately. Geometrically, we are saying that a unit sphere has curvature $R_{\rm sphere} = + 1$ and each polygon contributes to this curvature with $R_p = (6 - p) / 12$: a hexagon with $R_6 = 0$, a pentagon with $R_5 = + 1/12$, a heptagon with $R_7 = - 1/12$. The total energy is then the sum of bending energy and stretching energy, ${\cal E}_t = {\cal E}_b + {\cal E}_s$ that compete for the minimization of the total energy \cite{bnt}.

\subsection{Scar formation-annihilation mechanism for viruses}

\begin{figure}
\centering
\includegraphics[height=.2\textheight]{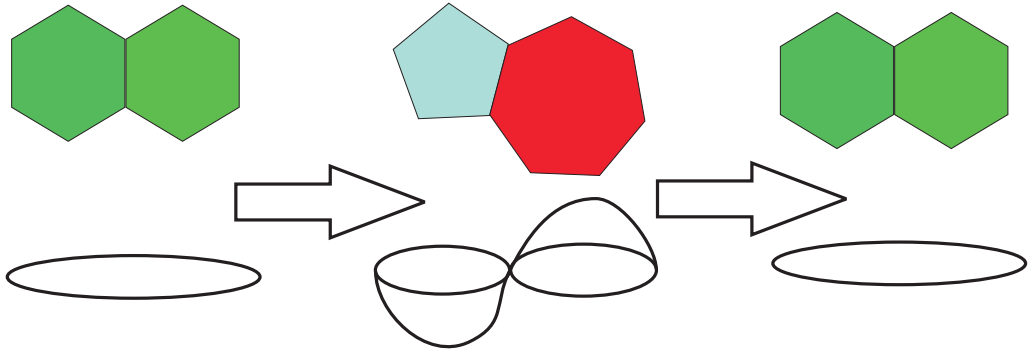}
  \caption{Idealized scar formation-annihilation mechanism.}
  \label{scar}
\end{figure}

Below the threshold for the scar production $N_{\rm scar}$, ${\cal E}_b$ is provided by the 12 pentagons, while the hexagons contribute to ${\cal E}_s$ only. One way to imagine the transition at $N > N_{\rm scar}$ is to think of a pair 6-6, with zero total and local curvature and zero bending energy, {\it converted into} a pair 5-7 (see Fig.~\ref{scar}), again with zero total curvature but with nonzero local curvature hence with nonzero bending energy given by $2 E_b$, where $E_b$ is the energy necessary to convert a 6 into a 5 or into a 7. In \cite{cejb2008} we proposed that scars can {\it appear on virus capsids} at an intermediate stage of their evolution towards a non-spherical shape and can actually \textit{drive} such shape-change. There $E_b$ should be related to the {\it conformational switching energy} \cite{speir} (for HIV-1 see, e.g., \cite{HIV-1CS}).

The (idealized) shape-change mechanism of \cite{cejb2008} is (see Fig.~\ref{scar}): i) proteins first make a CK icosahedron; ii) when, e.g., $N$ reaches $N_{\rm scar}$, they form scars: $6-6 \to 5-7$; iii) the capsid changes shape via the release of the bending energy into stretching energy at the location of the scar with the consequent ``annihilation'' of the 5-7 pair: $5-7 \to 6-6$. The resulting capsid has the CK morphological units but not the spherical shape.

\begin{figure}[h]
\begin{minipage}{14pc}
\includegraphics[height=.25\textheight, width=12pc]{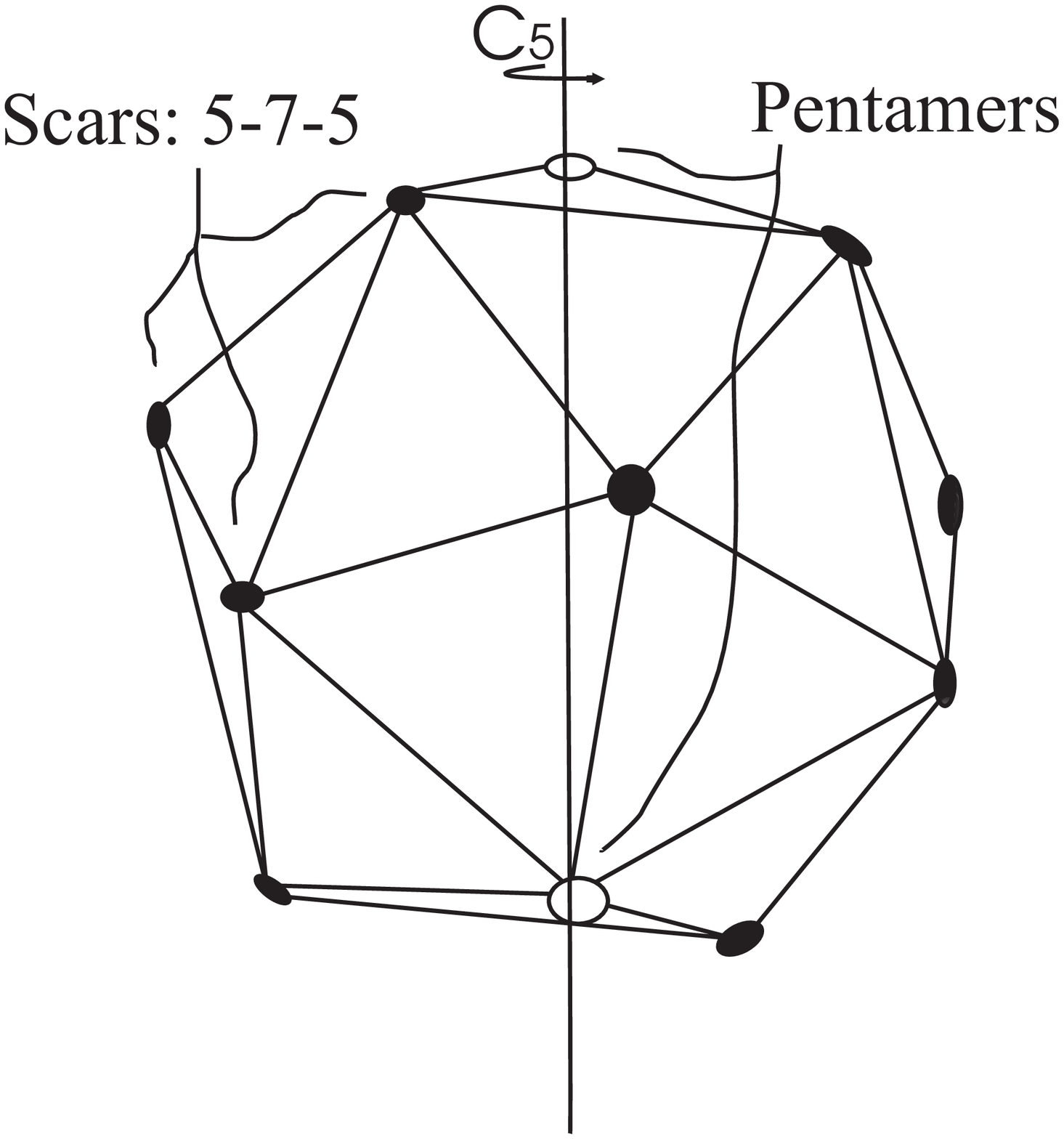}
\caption{The intermediate spherical capsid.}
\label{icoscar}
\end{minipage}\hspace{5pc}%
\begin{minipage}{14pc}
\includegraphics[height=.25\textheight, width=7pc]{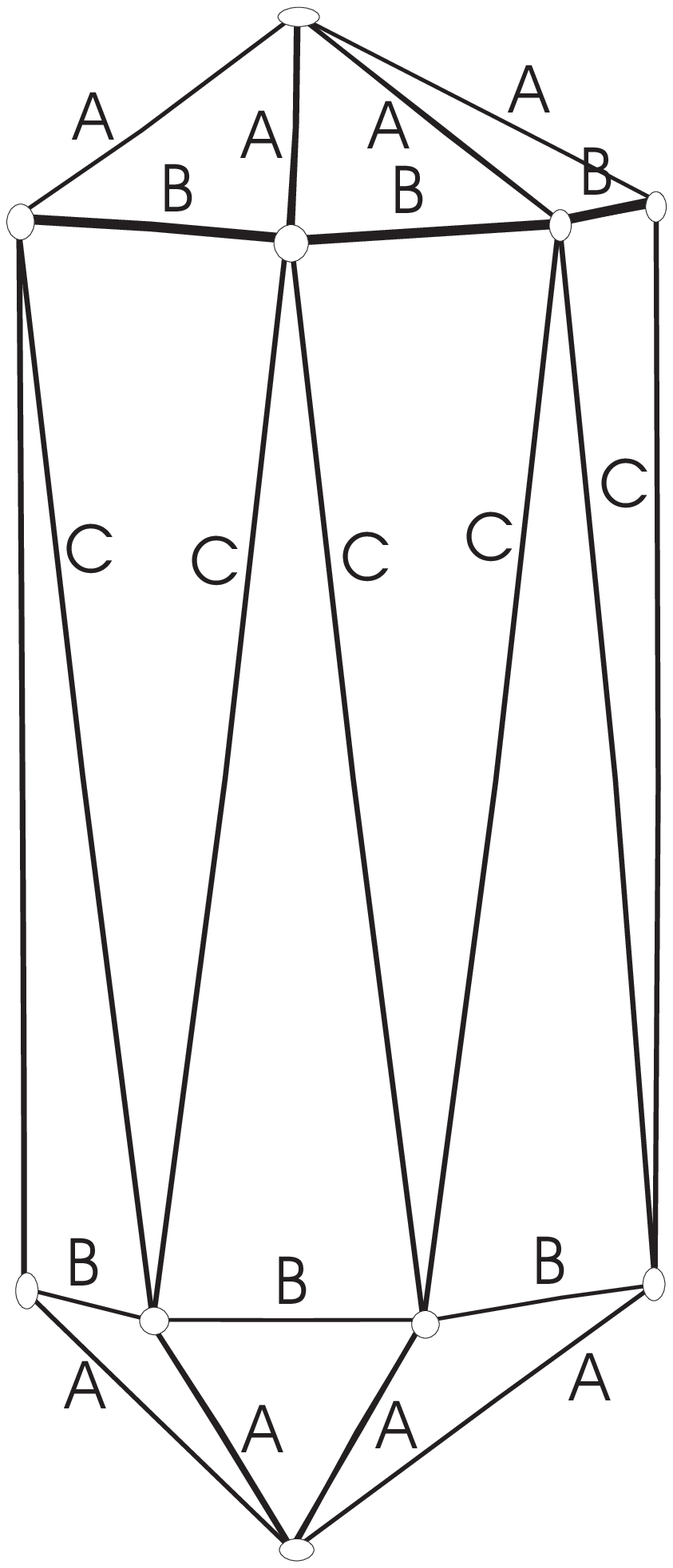}
\caption{The final spherocylindrical capsid.}
\end{minipage}
\end{figure}

An example given in \cite{cejb2008} is that of scars created only near the 10 inner vertices via a mechanism that respects a $C_5$ rotation symmetry around the north pole-south pole axis. In Fig.~\ref{icoscar} the vertices where the scars are formed are indicated with $\bullet$, while the other two are indicated with $\circ$. If we require that this mechanism is area preserving, i.e. the total number of proteins before and after is always $60T$ the final capsid obtained is the spherocylinder of Fig.~5 (a shape taken by certain bacteriophages' head) with the 12 pentamers at the vertices and the $10 (T - 1)$ hexamers distributed differently with respect to the intermediate icosadeltahedron. In general a variety of final capsid shapes could be obtained via this mechanism. For the $C_5$ symmetric example given, if, for instance, the scars carry a bigger bending energy the final shape could be a backy-tube capsid; if the orientation of the scars in the previous setting is such that $C$ shrinks, hence $B$ becomes longer (see Fig.~5) then a disk-like shape is obtained; etc..

\subsection{Are scars at work on the HIV-1 immature virions? The data of Sundquist's group.}
This is just an idealized model and to make full contact with real viruses one has to be ready to adjust it and, as scars are supposed to disappear,
it is also a delicate matter to find the right experimental set-ups for spotting the scars. A case that seems particularly promising is the important case (for human health) of the HIV-1. There the number of Gag polyproteins (essentially made of the three proteins MA, CA, NC) that make up the immature quasi-spherical capsid is estimated to be between 2500 \cite{sundquist2007} and 5000 \cite{briggs2004} (even 11000 is reported \cite{briggs2004}), i.e. well above the $N_{\rm scar}$ of the Thomson problem. Indeed, the impressive data of Sundquist's group taken by electron cryotomography of the immature capsid \cite{sundquist2007} (see also \cite{sundquistreview2008}) reveal the structure of the CA shell that is seen to have an hexameric texture and large areas with an apparent lack of symmetry (``disordered`` areas in the terminology of \cite{sundquist2007}). The role of these a-symmetric areas in the maturation process, the apparent lack of pentameric structures, the mechanism that allows for closure of the spherical capsid or the absence of such closure, and other aspects are important issues to be understood.

We claim here that in those areas of the capsid of the immature HIV-1 virions the scar formation-annihilation mechanism we discussed might be at work, although, since it takes place within a many-layer structure comprising the rest of the Gag and the lipidic envelop, it does not give the effect described above in the idealized case. What it presumably does instead is: (i) to stretch the structure, hence not allowing the perfect spherical shape; (ii) in combination with the proteolytic cleavage is responsible for the dramatic reduction of the number of proteins (the generally accepted number for the mature capsid is around 1500); (iii) it might even explain the apparent absence of pentamers if one supposes that when the annihilation takes places it involves also the ``primitive'' pentamers, i.e. what in the model are the untouched  (by the annihilation) vertices of the intermediate icosahedron.

Of course we need to explore those assertions in a quantitative fashion within the model, and we have work in progress in that direction. In the meanwhile, a faster way is to probe those claims directly in experiments, ideally with the help of Sundquist's laboratory.

\subsection{Effects of the lack of a general theory of biological physics}
Again in this case, as for the DNA compaction, it is clear that the appropriate general frame is lacking. With some adjustment (e.g., ``DNA packing'' goes to ``capsid shape-changes'', etc) all questions of Subsection~\ref{dnaquestions} apply here, except (i). One important point here is that viruses are at the {\it border} between dead matter (where standard physics applies) and living matter (where it appears that new physics is necessary), hence we deem them to be the best candidates to understand how to build-up the missing frame. For instance, invariably all models (including ours) on virus structure deal with (free) energy minimization (see \cite{bruinsmaRNAvirus} for a review), just like, for instance, an elastic ball. May be this is a place to start looking for changes.

\section{Biology as Physics}
So, what is it this new physics that we claim to be necessary for a true understanding of living matter? The vast majority among physicists and
biologists says that there is nothing to be done: the physics for living matter is just the same as for dead matter (elastic balls, charged wires,
classical machines at the nanometer scale, etc.) it is ``just more complicated to apply''. Even taking this conservative view, it seems to us that
(to say it with a metaphor) it is like trying to solve a (difficult) geometry problem with spherical symmetry insisting in using cartesian coordinates: results are more difficult to get, their interpretation becomes obscure and, most importantly, we might miss global information insisting with a description that is good locally but not globally (see also the metaphor of the mechanical engineer bumping into an electric motor of \cite{schroedinger}). The conservative approach to biophysics is nowadays achieving impressive results, and this approach is just what we have been using in our own work reviewed in the previous two Sections, but already at this level of the analysis (i.e. by just focusing on the problems to be solved and not on the general problems of method) we pointed out in Subsections 2.4 and 3.4 that the proper frame is missing and this is not just our own impression.

There are some calls to theoretical physics from the biology frontier to help systematize biology in a rigorous manner (see, e.g., \cite{biofront}). We physicists could partially answer these calls by, for instance, fitting a given biological problem into a given established physical theory and use the logical consistency of the latter. This is an exciting and worth thing to do and when it is possible to do that there is plenty of cross-fertilizations between the two fields. In a way our proposal of the scar creation-annihilation mechanism for viruses (and the whole ``relationship'' between the Thomson problem, or elastic theory of spherical membranes \cite{bnt} or the Landau-Ginzburg theory of phase transitions \cite{bruinsmaguerin} on the physics side and virus structure on the biology side) is an example. Then the application of physical theories to biological problems might make {\it that} sector of biology just like {\it that} physical theory, hence the frame in that case would be set without mayor changes in the methods. This, though, would not solve the problem in general but would work only for certain particular features of certain particular biological entities and it is not certain at all that this would work all the time (and why should it?).

We have now identified what is missing: a general theory of physical biology (GTPB), namely, {\it the} conceptual frame that should {\it always} work, for all biological problems, no matter the complexity, and that ``just'' needs to be applied case by case. It must give the correct description of the phenomenon and, via mathematical elaborations, must produce predictions. This is surely a tremendous task. Let us now see whether this is  accomplishable with building blocks we have already or whether we need to change something on the physics side as well.

In 1943 such a change of physics to move the frontier of knowledge from dead to living matter was deemed necessary by Schr\"{o}dinger
in his famous lectures delivered at Trinity College in Dublin \cite{schroedinger}. Two were the main messages there: the carrier of genetic information needs to be an aperiodic solid and the thermodynamics of living matter is based on the order from order paradigm rather than the order from disorder
of dead matter. Both considerations were rooted into quantum mechanics.

The first message was a clear prediction of the structure and role of DNA, discovered only ten years later. The second message is still an open question that found some partial answers in the work of Prigogine and others who invented Non-Equilibrium Thermodynamics (NET) (see, e.g., \cite{volkenstein} for a textbook chapter). Since living organisms are far from equilibrium open systems, NET is surely one of the building blocks we need for GTPB, but it cannot be the end of the story because, similar considerations apply to other non-living systems (financial markets, traffic jams, weather systems, etc, i.e. what is sometimes called a ``complex system'') and this is not what we are looking for. We need a tailored physics for life, or, in other words, what we might call a ``principle of distinction'' at work {\it exclusively} for living matter, and this is still not there. Another point is that the role of quantum mechanics as pivotal for life phenomena \cite{schroedinger, penrose} needs to be fully exploited.

To build up the GTPB is the most exciting and challenging problem ahead, but it is very difficult. For the time being, we can solve the many challenging problems of standard biophysics, investigate better the role of quantum physics for life and build up a ``GTPV'': a general theory of physics for viruses.

\ack It is my pleasure to thank Ji\v{r}i Horej\v{s}i and the Institute of Particle and Nuclear Physics of Charles University for supporting my incursions into biological physics, Samik Sen for landing his own time to the DNA project and, finally, Siddhartha Sen for joyfully joining in from the very beginning.

\end{document}